\documentclass[aps,pr,twocolumn,groupedaddress,notitlepage,
floatfix,superscriptaddress]{revtex4-2}

\pdfoutput=1 
\usepackage{graphicx,graphics,epsfig,subfigure,times,bm,bbm,amssymb,amsmath,amsthm,mathrsfs,MnSymbol} \usepackage{gensymb} \usepackage{amsfonts} \usepackage{float} \usepackage[matrix,frame,arrow]{xypic} \usepackage[pdfstartview=FitH]{hyperref} 
\usepackage{times}
\usepackage{float}
\usepackage{graphics}
\usepackage[T1]{fontenc}

\usepackage{braket}  
\usepackage{enumerate} \usepackage[normalem]{ulem}
 \usepackage[usenames,dvipsnames]{xcolor} \usepackage{multirow} \usepackage{mathtools}
 \usepackage{bbm} \usepackage{titletoc} 

 \definecolor{orange}{rgb}{1,0.5,0} 

 \newcommand{\RNum}[1]{\uppercase\expandafter{\romannumeral #1\relax}}

 \newcommand{\ignore}[1]{}
 \usepackage{geometry}

 \ignore{ \documentclass[eprintnumbers,amsmath,amssymb,onecolumn,a4paper,caption ]{article}
 \usepackage{amsfonts} \usepackage{amssymb} \usepackage{mathrsfs} \usepackage{mathbbold}
 \usepackage{bbm} \usepackage{mathrsfs} 
 \usepackage{dcolumn}
 \usepackage{bm}
 \usepackage{times,epsfig,amssymb,amsmath} \usepackage{float} \usepackage{subfigure}
 \usepackage{geometry}\geometry{left=2.5cm,right=2.5cm,top=3cm,bottom=3cm}
   \usepackage{color}

   }
\hypersetup{ colorlinks=true,       
	linkcolor=red,          
	citecolor=blue,        
	filecolor=magenta,      
	urlcolor=blue,           
	runcolor=cyan }

\begin{document}
	\title{Observation of Emergent  $\mathbb{Z}_2$  Gauge Invariance in a Superconducting Circuit}
	
	\author{Zhan Wang}
	\altaffiliation[]{These authors contributed equally to this work.}
	\affiliation{Beijing National Laboratory for Condensed Matter Physics, Institute of Physics, Chinese Academy of Sciences, Beijing 100190, China}
	\affiliation{School of Physical Sciences, University of Chinese Academy of Sciences, Beijing 100190, China}
	
	\author{Zi-Yong~Ge}
	\altaffiliation[]{These authors contributed equally to this work.}
	\affiliation{Beijing National Laboratory for Condensed Matter Physics, Institute of Physics, Chinese Academy of Sciences, Beijing 100190, China}
	\affiliation{School of Physical Sciences, University of Chinese Academy of Sciences, Beijing 100190, China}
	
	\author{Zhongcheng~Xiang}
	\altaffiliation[]{These authors contributed equally to this work.}
	\affiliation{Beijing National Laboratory for Condensed Matter Physics, Institute of Physics, Chinese Academy of Sciences, Beijing 100190, China}

	\author{Xiaohui~Song}
\affiliation{Beijing National Laboratory for Condensed Matter Physics, Institute of Physics, Chinese Academy of Sciences, Beijing 100190, China}

	\author{Rui-Zhen Huang}
\affiliation{Kavli Institute for Theoretical Sciences, University of Chinese Academy of Sciences, Beijing 100190, China}

	\author{Pengtao~Song}
    \affiliation{Beijing National Laboratory for Condensed Matter Physics, Institute of Physics, Chinese Academy of Sciences, Beijing 100190, China}
	\affiliation{School of Physical Sciences, University of Chinese Academy of Sciences, Beijing 100190, China}

	\author{Xue-Yi~Guo}
\affiliation{Beijing National Laboratory for Condensed Matter Physics, Institute of Physics, Chinese Academy of Sciences, Beijing 100190, China}

	\author{Luhong~Su}
\affiliation{Beijing National Laboratory for Condensed Matter Physics, Institute of Physics, Chinese Academy of Sciences, Beijing 100190, China}
\affiliation{School of Physical Sciences, University of Chinese Academy of Sciences, Beijing 100190, China}

	\author{Kai~Xu}
	\affiliation{Beijing National Laboratory for Condensed Matter Physics, Institute of Physics, Chinese Academy of Sciences, Beijing 100190, China}
	\affiliation{Songshan Lake Materials Laboratory, Dongguan 523808, Guangdong, China}
	\affiliation{CAS Center for Excellence in Topological Quantum Computation, UCAS, Beijing 100190, China}
	
	\author{Dongning~Zheng}
	\email{dzheng@iphy.ac.cn}
	\affiliation{Beijing National Laboratory for Condensed Matter Physics, Institute of Physics, Chinese Academy of Sciences, Beijing 100190, China}
	\affiliation{School of Physical Sciences, University of Chinese Academy of Sciences, Beijing 100190, China}
	\affiliation{Songshan Lake Materials Laboratory, Dongguan 523808, Guangdong, China}
	\affiliation{CAS Center for Excellence in Topological Quantum Computation, UCAS, Beijing 100190, China}
	
	\author{Heng~Fan}
	\email{hfan@iphy.ac.cn}
	\affiliation{Beijing National Laboratory for Condensed Matter Physics, Institute of Physics, Chinese Academy of Sciences, Beijing 100190, China}	
	\affiliation{Songshan Lake Materials Laboratory, Dongguan 523808, Guangdong, China}
	\affiliation{CAS Center for Excellence in Topological Quantum Computation, UCAS, Beijing 100190, China}
	\affiliation{Beijing Academy of Quantum Information Sciences, Beijing 100193, China}

	\begin{abstract}
		Lattice gauge theories (LGTs) are one of the most fundamental subjects in  many-body physics,
		and has recently attracted considerable research interests in quantum simulations.
		Here we experimentally investigate the emergent $\mathbb{Z}_2$ gauge invariance in a 1D superconducting circuit with 10 transmon qubits.
		By precisely adjusting staggered longitudinal and transverse fields to each qubit,
		we construct an effective Hamiltonian containing an LGT and gauge-broken terms.
		The corresponding matter sector can exhibit a localization,
		and there also exists  a 3-qubit operator, of which the expectation value can retain nonzero for a long time in low-energy regimes.
		The above localization can be regarded as the confinement of matter fields, and the 3-body operator is the $\mathbb{Z}_2$ gauge generator.
		These experimental results demonstrate that, despite the absence of gauge structure in the effective Hamiltonian, $\mathbb{Z}_2$ gauge invariance can still emerge in  low-energy regimes. 
		Our work provides a method for both theoretically and experimentally studying the rich physics in quantum many-body systems with emergent gauge invariance.
	\end{abstract}

	\maketitle
	
	\textit{Introduction.}---%
	Gauge invariance is one of the most fundamental principles of quantum field theories, 
	and lattice gauge theories (LGTs)~\cite{Wilson1974,Kogut1979,Wen2004,Fradkin2013} play a significant role in a wide range of modern physics,
    e.g.,  helping us understand the confinement of quarks~\cite{Wilson1974}. 
	Recently, as the rapid development of quantum simulations~\cite{Buluta2009,Georgescu2014},
	studying LGTs in synthetic quantum many-body systems becomes possible 
	and has drawn many interests from both theoretical and experimental physicists~\cite{Zohar2012,Banerjee2012,Barbiero2019,Hauke2017,Marcos2013,Brennen2016,Zohar2017}.
	On the one hand, many unresolved problems associated with LGTs are potentially solvable via large-scale quantum simulations.
	On the other hand, quantum simulations provide a new viewpoint to study LGTs, i.e., nonequilibrium dynamics~\cite{Polkovnikov2011,Eisert2015,Hebenstreit2013,Kormos2017}.
	The corresponding experimental studies have been demonstrated in ultracold atoms~\cite{Schweizer2019,Yang2019,Gorg2019}.

	In condensed matter physics,  gauge invariance is generally not a necessary element and thus absent in the Hamiltonian.
	However, due to strong quantum fluctuations of quantum many-body systems,
    gauge invariance may emerge in  low-energy regimes  leading to various  novel quantum phases, such as quantum spin liquid~\cite{Kitaev2003_1,Kitaev2003_2,Zhou2016} and deconfined  quantum critical points~\cite{Senthil2004}.
	In addition,  it is challenging to construct an exact gauge-invariant Hamiltonian in an artificial quantum system~\cite{Barbiero2019,Schweizer2019}.
	Thus one natural question is whether we can construct a quantum many-body system that the gauge structure is absent in the Hamiltonian but emergent in the low-energy physics.
	Moreover, this emergent gauge invariance may lead to novel dynamics, for instance, the confinement induced localization of matter degrees of freedom~\cite{Kormos2017,Schweizer2019},
	which can be detected by state-of-the-art quantum simulators.

		\begin{figure*}[t]     		
		\includegraphics[width=0.95\textwidth]{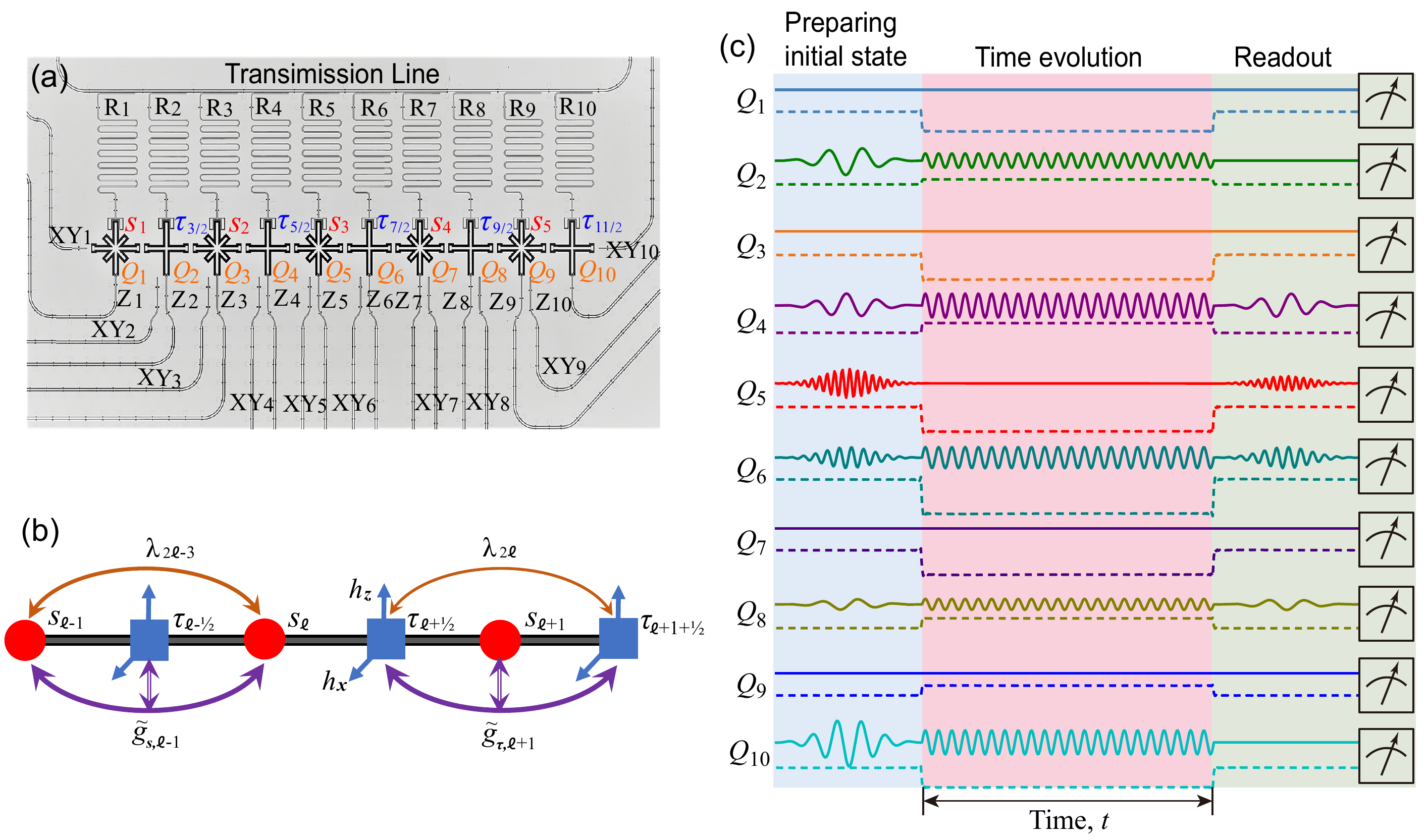}
		\caption{Set-up and protocol. (a)  Optical micrograph of the  superconducting
			circuit. There are 10 transmon qubits arranged into a chain,  where the odd and even qubits are used for realizing $s$- and $\tau$-spins in the following experiment, respectively.  Each qubit couples to a microwave line for the $XY$ driving ($\text{XY}_j$), a flux bias line for the $Z$ pulse ($\text{Z}_j$), and
			a readout resonator for measurement ($\text{R}_j$). 
			The NN qubits are directly coupled via a capacitor, and the capacitance of each qubit itself contributes to the NNN coupling.
			The parameters of the device are presented in the Supplemental Material (SM) in detail~\cite{SM}.		
			(b) Lattice skeleton of the effective Hamiltonian~(\ref{Heff}).  There are two types of spins, i.e., $s$ (red sites) and $\tau$ (blue sites), 
			corresponding to matter and gauge fields, respectively. The purple arrows represent the three-body couplings,  the orange arrows are direct NNN two-body couplings, and the blue arrows are  longitudinal/transverse fields. (c) Pulse sequences of the experiment. The solid and dashed lines represent XY drivings and Z pulses, respectively.
			First, all qubits are at their idle frequencies for preparing the initial state $\ket{\psi_0}$ via XY driving. Then, each qubit is biased to the working point with $Z$ pulses, and meanwhile, the XY driving  is applied to each even qubit. After the system evolves for time $t$, 
			all qubits are biased back to idle points for readout.}
		\label{fig_1}
	\end{figure*}
	
	In this Letter, we experimentally demonstrate the emergent $\mathbb{Z}_2$
	gauge invariance in a  superconducting processor with 10 qubits arranged into a chain.
	Following the scheme in Ref.~\cite{Ge2020}, we apply stagger longitudinal and transverse fields to the system to construct an effective Hamiltonian,
	which is a mixture of a $\mathbb{Z}_2$ LGT and some gauge-broken terms.
	To detect the emergent $\mathbb{Z}_2$ gauge invariance, we first study the charge spreading of the matter sector.
	The experimental result shows that it can exhibit a localization under  proper parameters,
	which is a strong dynamical signature of the confinement induced by emergent gauge invariance.
	Then, benefiting from the joint readout of three qubits under the arbitrary basis, we can study the time evolution of 3-qubit operators. We find that there is a 3-qubit operator, whose expectation value can retain  nonzero for a long time. 
	We demonstrate that this operator can be regarded as the $\mathbb{Z}_2$ gauge generator.
	Therefore this experimental result is a direct evidence of the emergent $\mathbb{Z}_2$ gauge invariance. Furthermore, according to relations between charge spreading  and initial states, we find that the gauge structure mainly emerges in low-energy regimes.

	\textit{Model and set-up.}---%
	Due to the scalability, long coherent time, and high-precision full control, superconducting circuits~\cite{Makhlin2001,Gu2017} become one of the most competitive candidates for achieving the universal quantum computation~\cite{Arute2019,Wu2021},
	and they are also a more suitable platform for performing quantum simulations~\cite{Xu2018,Roushan2017,Salathe2015,Barends2015,Zhong,Song1,Flurin2017,Ma2019,Yan2019,Ye2019,Guo2019,Xu2020,Guo2021,Guoxy2021}. 
	Several theoretical schemes have been  proposed to simulate LGTs via specific superconducting circuits~\cite{Marcos2013,Brennen2016}.
	However, the corresponding experimental realization is still absent, which may be due to the technological challenges for preparing such superconducting circuits.
     Here inspired by the scheme in Ref.~\cite{Ge2020}, we find that the realization of LGTs is feasible by using conventional chain like transmon qubits.

	This experiment is performed in a superconducting circuit with 10 transmon qubits ($Q_1$--$Q_{10}$) arranged into a  chain, see Fig.~\ref{fig_1}(a).
	Due to the large and staggered anharmonicity~\cite{SM}, the system can be described by an isotropic 1D $XY$ model with tunable transverse and longitudinal fields~\cite{Schweizer2019,Yang2019,Gorg2019}.
	In addition, the next-nearest neighbor (NNN) coupling cannot be neglected in the following experiment.
	Therefore the Hamiltonian can be written as
	\begin{align} \label{Hxy} \nonumber
		\hat H =&\sum_{j}(g_{j}\hat{\sigma}^+_{j} \hat{\sigma}^-_{j+1}+\lambda_j\hat{\sigma}^+_{j} \hat{\sigma}^-_{j+2}+\text{H.c.})\\
		&+\sum_{j=1}(\frac{V_{ j}}{2}\hat{\sigma}^z_{j}+ h_{ x,j}\hat{\sigma}_{j}^x),
	\end{align}
	where $\hat{\sigma}^{\pm} = (\hat{\sigma}^x\pm i\hat{\sigma}^y)/2$, $\hat{\sigma}^{x,y,z}$
	are Pauli matrices, $g_{j}/2\pi\approx12$~MHz
	and $\lambda_{j}/2\pi\approx 0.7-1.1$~MHz are the nearest-neighbor (NN) and NNN coupling strengths~\cite{SM}, respectively,
	$V_j$ is a longitudinal field tuned by Z pulses, and $h_{x,j}$ is a transverse field controlled by XY drivings.

	Here we let the longitudinal field at each odd qubit $V_{2\ell-1}/2\pi=-80$~MHz
	and  the detuning between NN qubits is much larger than the coupling strength, i.e., $V_{2\ell}-V_{2\ell-1}\gg g_j$.
	In addition, the transverse field is only applied at the even qubits with equal strength $h_x$.
	Thus  according to Ref.~\cite{Ge2020}, we can obtain an effective Hamiltonian $\hat H_{\text{eff}}$, which reads
	\begin{align} \label{Heff}
		& \hat  H_{\text{eff}}=\hat H_1 + \hat H_2+\hat H_3, \\ \nonumber
		&\hat  H_1 =\sum_{\ell=1}^{4}\tilde{g}_{s,\ell}(\hat{s}^+_{\ell} \hat{\tau}^z_{\ell+\frac{1}{2}}\hat{s}^-_{\ell+1} +
		\text{H.c.} ) +\sum_{\ell=1}^{5}h_x\hat{\tau}_{\ell+\frac{1}{2}}^x  ,\\ \nonumber
		& \hat H_2 = -\sum_{\ell=1}^{4}\tilde{g}_{\tau,\ell}(\hat{\tau}^+_{\ell-\frac{1}{2}} \hat{s}^z_{\ell}\hat{\tau}^-_{\ell+\frac{1}{2}}+ \text{H.c.}),\\ \nonumber
		& \hat H_3 \!= \!\sum_{\ell=1}^{4}(\lambda_{2\ell-1}\hat{s}^+_{\ell} \hat{s}^-_{\ell+1} \!+\!\lambda_{2\ell}\hat{\tau}^+_{\ell+\frac{1}{2}} \hat{\tau}^-_{\ell+\frac{3}{2} }+
		\text{H.c.} ) \!+\!\sum_{\ell=1}^{5}h_z\hat{\tau}_{\ell+\frac{1}{2}}^z ,
	\end{align}
	where $\hat s_\ell = \hat \sigma_{2\ell-1}$ and $\hat\tau_{\ell+\frac{1}{2}}= \hat \sigma_{2\ell}$ are also Pauli matrices labeling the odd and even qubits, respectively, and $\tilde{g}_{s/\tau,\ell}\approx g_j^2/\Delta\approx-2\pi\times1.8$~MHz is the effective three-body coupling strength, see Fig.~\ref{fig_1}(b). Here the effective  longitudinal field $h_z$ is the summation of the original longitudinal field of even qubits ($V_{2\ell}$) and the Lamb shift~\cite{SM}.
	In the experiment, $V_{2\ell}$ is adjustable by controlling the detuning between qubit frequencies and XY driving frequencies.

	Here $\hat H_1$ is nothing but a $\mathbb{Z}_2$ lattice gauge field ($\tau$-spins) coupled to a matter field ($s$-spins),
	where the corresponding $\mathbb{Z}_2$  gauge generator reads $\hat{G}^0_\ell = \hat \tau^x_{\ell-\frac{1}{2}}\hat s^z _\ell\hat \tau^x_{\ell+\frac{1}{2}}$, i.e., $[\hat G_\ell^0, \hat H_1]=0$~\cite{Borla2020}.
	However, we can find that additional terms $\hat H_2$  and $\hat H_3$ both violate this gauge invariance,
	so the whole Hamiltonian is not a rigorous LGT.
	However, by numerical simulation via density matrix renormalization group (DMRG) method~\cite{Schollwock2005,Schollwock2011} (the details are shown in the SM~\cite{SM}), 
	we find that a new $\mathbb{Z}_2$ gauge invariance can still emerge in the ground state for a large transverse field $h_x$ and proper longitudinal field $h_z$.
	The corresponding new gauge generator  is  
	\begin{align} \label{Gl}
		\hat{G}_\ell= {\tilde\tau}_{\ell-\frac{1}{2}}^x\hat{s}_\ell^z{\tilde\tau}_{\ell+\frac{1}{2}}^x,
	\end{align}
	where $\tilde\tau^x=\sin \beta \hat\tau^z + \cos\beta\hat\tau^x$ and $\beta= \arctan (h_z/h_x)$.

	Next, we will experimentally investigate quench dynamics of this system to probe the emergent gauge invariance.
	 Here  the gauge invariance is almost independent of the filling of $s$-spins~\cite{SM}.
	 Thus  without loss of generality,
	we consider the system containing only one $s$-spin, i.e., $\sum_{\ell=1}^5 \hat{s}_\ell^+ \hat s_\ell^- = 1$.
	The initial state is chosen as $|\psi_0\rangle = |s\rangle\otimes|\tau\rangle$, 
	where  $|s\rangle$ and $|\tau\rangle$ label the states of $s$- and $\tau$-spins, respectively.
	We let $|s\rangle = \ket{00100}$ and $|\tau\rangle=\ket{\Phi_\theta\Phi_\theta\Phi_\theta\Phi_\theta\Phi_\theta}$, where$\ket{\Phi_\theta}=\cos \frac{\theta}{2}\ket{1}+\sin\frac{\theta}{2}\ket{0}$.
	Here  according to the effective Hamiltonian $\hat H_{\text{eff}}$, we know that different $\theta$ means different energy  during the quench dynamics.
	The experimental procedure can be summarized as follows: First, all qubits are at their idle frequencies, and we use single-qubit rotational gates to prepare the initial state $\ket{\psi_0}$. Then, we bias the frequency of each qubit to the corresponding working point with Z pulses, i.e., let the local potential of each qubit be $V_j$.
	Meanwhile, we apply ac drivings with amplitude $h_x$ to even qubits through  XY lines to realize transverse fields.
	Finally, after the system evolves with time $t$, we bias back all qubits to their idle points, and read out the corresponding observable.
	The pulse sequence of each qubit is shown in Fig.~\ref{fig_1}(c).

	\begin{figure}[t]     		
		\includegraphics[width=0.48\textwidth]{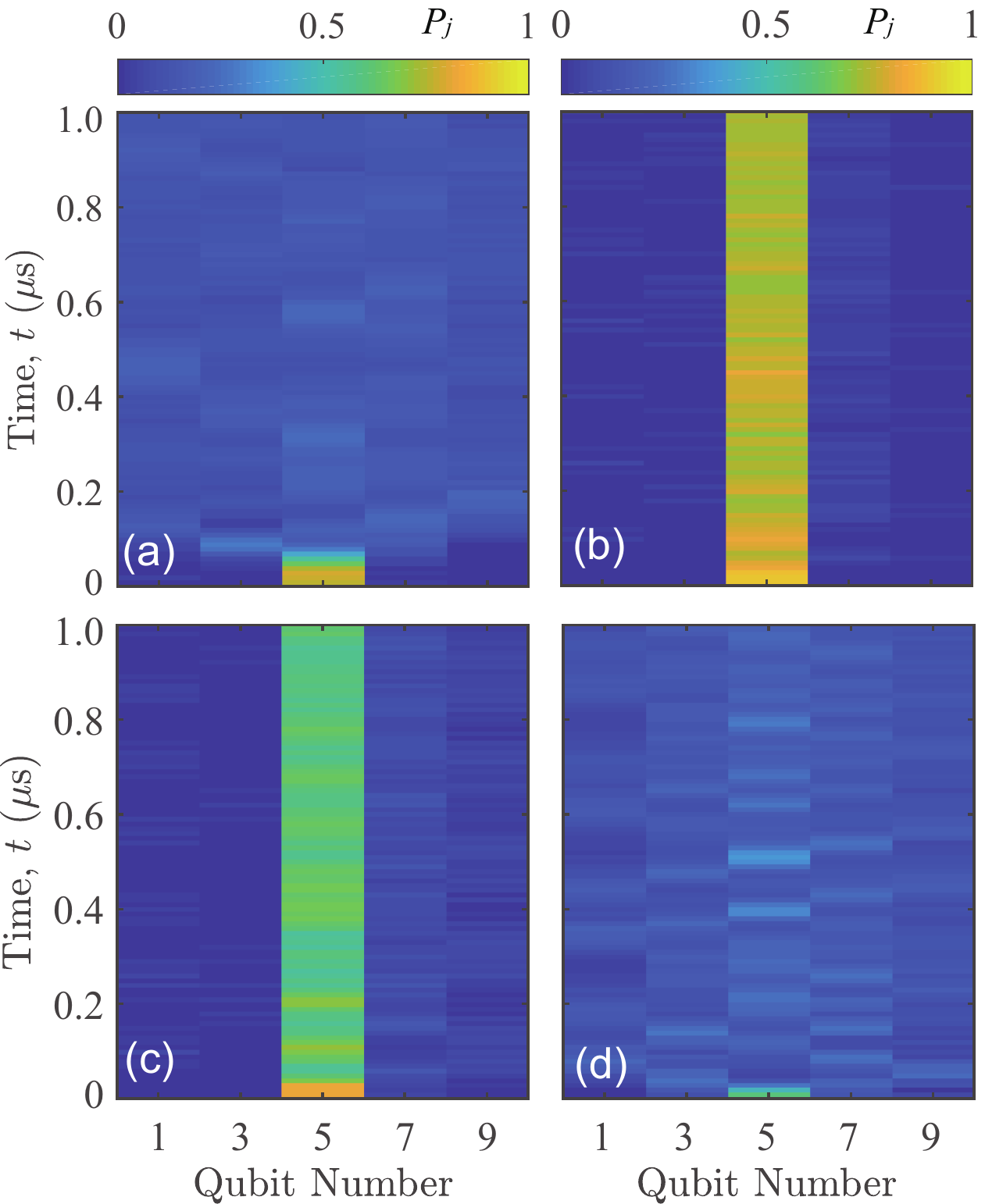}
		\caption{Spreading of $s$ spins up to $1$~$\mu$s for (a)  $h_x/2\pi=2$~MHz, $V_{2\ell}/2\pi=0$~MHz, $\theta =- \pi/3$;
			(b)  $h_x/2\pi=6$~MHz, $V_{2\ell}/2\pi=15$~MHz,  $\theta = -\pi/3$; (c) $h_x/2\pi=6$~MHz, $V_{2\ell}/2\pi=15$~MHz,  $\theta = -\pi/2$;
			and (d) $h_x/2\pi=6$~MHz, $V_{2\ell}/2\pi=15$~MHz, $\theta = \pi$. 
			Here  the $s$-spin shows localization in the cases of (b--c) indicating the presence of confinement, 
			while it transports freely in (a) and (d) indicating the absence of confinement.
			Each point is the average of $8,000$ single-shot readouts.}
		\label{fig_2}
	\end{figure}

	\textit{Confinement dynamics.}---%
	In 1D $\mathbb{Z}_2$ LGTs, the system is generally in a confined phase.
	In this case,  similar to the confinement of quarks in particle physics, the presence of a gauge field will induce a long-range potential of matter fields, e.g., linear potential~\cite{Surace2020}.
	This long-range potential can suppress the transport of matter field and results in a localization~\cite{Borla2020,Schweizer2019}.
	Thus   we can  study the dynamics of $s$-spins to demonstrate whether they can localize, which is a signature for identifying the existence of emergent  gauge invariance.
	Here we measure the density distributions of the photon at odd qubits, i.e., the spin-density distribution of the $s$ sector, defined as 
	\begin{align} \label{Pj}
	P_j(t) := \bra{\psi(t)} \hat{\sigma}^+_j\hat{\sigma}^-_j\ket{\psi(t)},
	\end{align}
	where $\ket{\psi(t)} = \mathrm{\exp}(-i\hat{H} t) \vert \psi_0 \rangle$ is the wave function of the system at time $t$.
	
	In the case of $h_x/2\pi=2$~MHz, $V_{2\ell}/2\pi=0$~MHz, and  $\theta = -\pi/3$,
	we can find that  $s$-spins  delocalize and spread to the whole system very quickly, see Fig.~\ref{fig_2}(a).
	However, the situation is different when $h_x/2\pi=6$~MHz, $V_{2\ell}/2\pi=15$~MHz.
	Figs.~\ref{fig_2}(b--c) show that $s$ spin can indeed localize when $\theta = -\pi/3$ or $-\pi/2$,
	which is an evidence of existing confinement.
	When we continue to change the initial state, e.g., $\theta=\pi$, the localization of $s$ spins disappears, see Fig.~\ref{fig_2}(d).
	These experimental results indicate that  the confinement of
	$s$ spins can indeed emerge for the proper transverse and longitudinal fields, and also depend on the system energy.

	\begin{figure}[t]     		
		\includegraphics[width=0.43\textwidth]{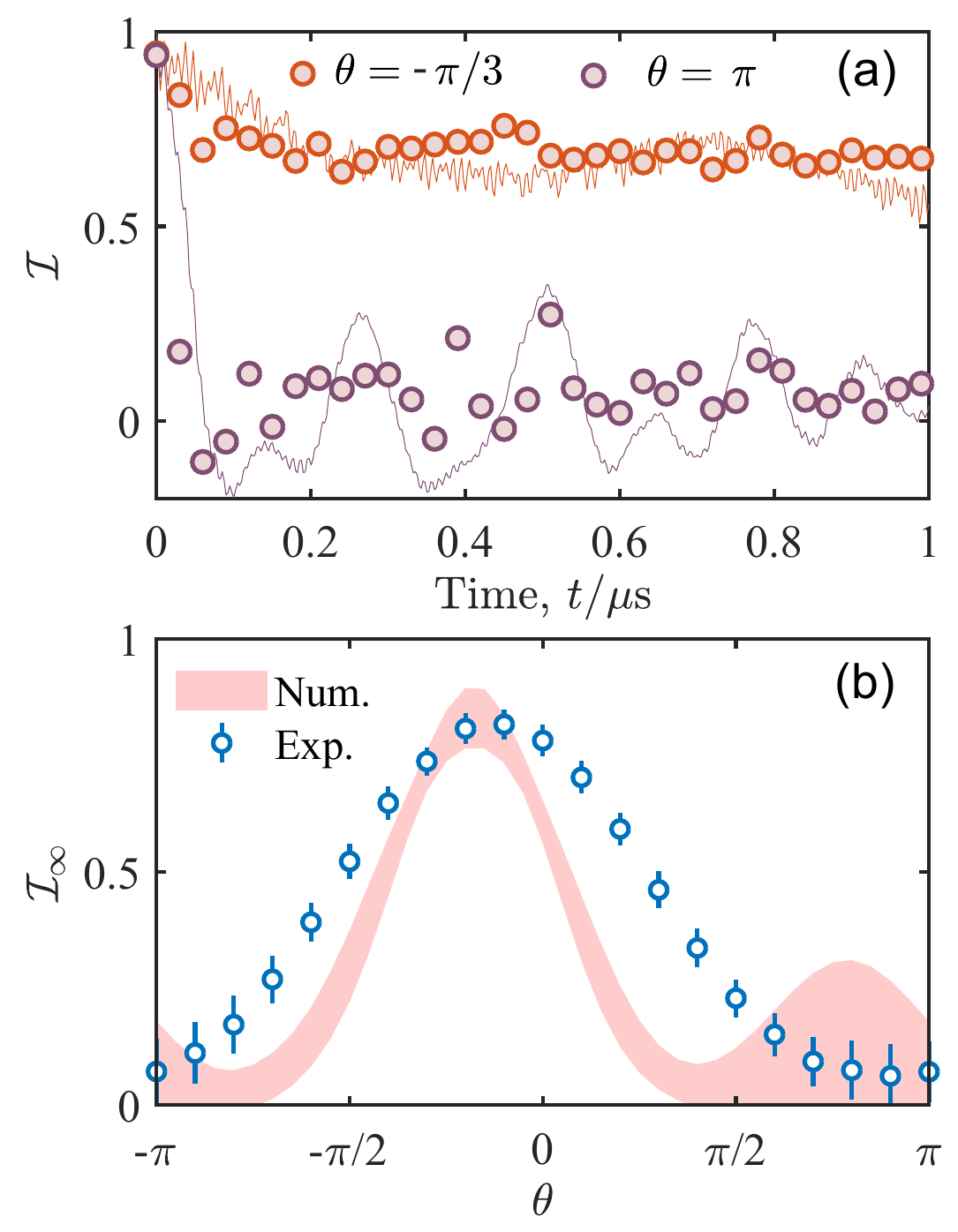}
		\caption{Extended imbalance $\mathcal{I}$ versus initial states. (a) Time evolution of  $\mathcal{I}$ with $h_x/2\pi=6$~MHz, $V_{2\ell}/2\pi=15$~MHz and different $\theta$,
			i.e., cases in Figs.~\ref{fig_2}(b--c). 
			The dots are experimental data, while the solid curves are the corresponding numerical results.
			(b) The relation between $\mathcal{I}_\infty $ and $\theta$. 
			Here, $\mathcal{I}_\infty $  is calculated as the average of $\mathcal{I}$ from 0.2 to 1~$\mu$s, and the standard deviation is the corresponding variance.
			Using Gaussian fitting, we obtain the peak of experimental result is at $\theta=\theta_m \approx - 0.35$.
			The numerical results are obtained via Hamiltonian in Eq.~(\ref{Hxy}), where  decoherence and dephasing are both neglected. }
		\label{fig_3}
	\end{figure}
	
	Next, we further explore the relation between confinement  and the system energy,
	which can be characterized  by the localization strength of $s$ spins.
	Here we define the extended imbalance of $s$ spins~\cite{Guo2021}
	\begin{align} \label{Iex}
		\mathcal{I}:= \sum_{j=\text{odd}}\eta_j P_j,
	\end{align}
	where $\eta_j=1/N_1\ (-1/N_0)$ if the initial state of $Q_j$ is $\ket{1}\ (\ket{0})$,
	and $N_1\ (N_0)$ is the number of $\ket{1}\ (\ket{0})$ for the initial state of the $s$-sector.
	In Fig.~\ref{fig_3}(a), we show the dynamics of extended imbalance for different $\theta$ under the condition of $h_x/2\pi=6$~MHz and $V_{2\ell}/2\pi=15$~MHz.
	It shows that $\mathcal{I}$ can stabilize at different values indicating the different localization strength.
	Now we use the steady value of extended imbalance $\mathcal{I}_\infty$  to quantify the localization strength,
	which can be calculated as the average of $\mathcal{I}$ during the last 0.8~$\mu$s.
	Here the larger $\mathcal{I}_\infty $ means the stronger localization strength.
	In Fig.~\ref{fig_3}(b),  the relation between $\mathcal{I}_\infty $ and $\theta$ is presented.
	We can find that, when $\theta=\theta_m \approx - 0.35$,
	the corresponding $\mathcal{I}_\infty $ is the largest indicating the strongest localization strength in this case.
	In the following discussion, we will verify that this is because the emergent gauge invariance can only exist in a low-energy regime, 
	and the initial state is close to the ground state when $\theta=\theta_m$.

	\begin{figure}[t]     		
	\includegraphics[width=0.4\textwidth]{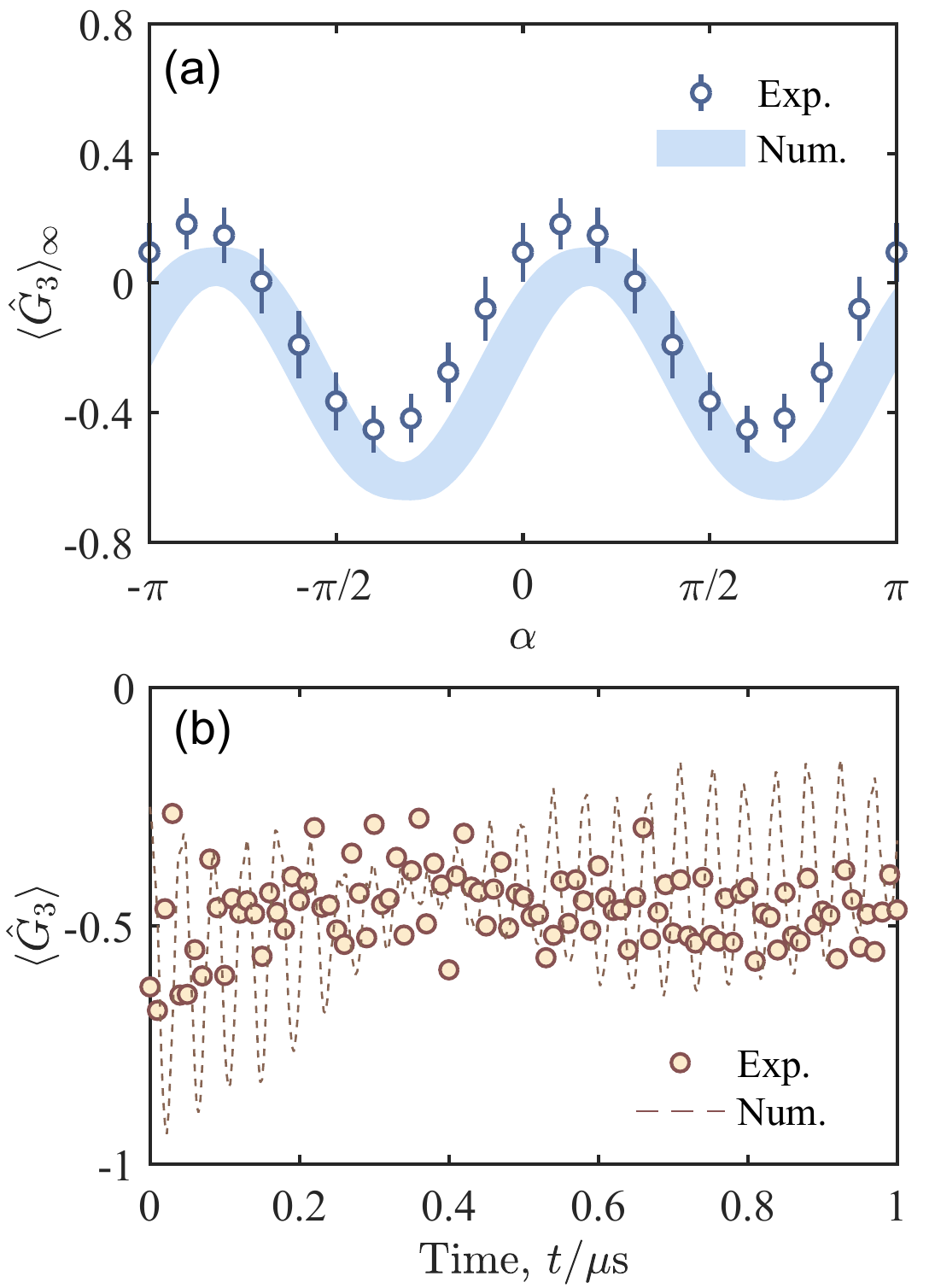}
	\caption{Gauge invariance. (a) The relation between $\braket{\hat G_3}_\infty$ and $\alpha$ with the parameters being identical to the cases in Figs.~\ref{fig_2}(b).
		Here, $\braket{\hat G_3}_\infty$, calculated as the average of $\mathcal{I}$ from 0.2 to 1~$\mu$s with the standard deviation being the error bar,
		represents the steady expectation value of 3-qubit operator $\hat{G}_3 (\alpha)$, defined in Eq.~(\ref{Qj}). 
		The curve is almost a sine-shape curve with the valley at $\alpha\approx -1.19$.
		(b) Time evolution of 3-qubit operator $\braket{\hat G_3(\alpha=-1.19)}$ up to 1~$\mu$s. 
	    It shows that $\braket{\hat G_3}$ can retain a nonzero value for long time, where the oscillation is from the high-order term~\cite{SM}.} 
	\label{fig_4}
\end{figure}
	
	\textit{Gauge invariance.}---%
	Now we start to directly study the gauge invariance of the system.
	Here  when deriving the effective Hamiltonian $\hat H_{\text{eff}}$,
	the correction of longitudinal field $h_z$ from high-order terms can hardly be directly confirmed accurately.
	Thus the $\beta$ or $\mathbb{Z}_2$ gauge generator $\hat{G}_\ell$  is in fact unknown.
	However,  we can define an ansatz of $\mathbb{Z}_2$ gauge generator as
	\begin{align} \label{Qj}
		\hat{G}_\ell (\alpha) :=\hat T_{\ell-\frac{1}{2}}(\alpha)\hat{s}_\ell^z \hat T_{\ell+\frac{1}{2}}(\alpha),
	\end{align}
	where $\hat T_{\ell-\frac{1}{2}}(\alpha) = \cos (\alpha)\hat\tau^x_{\ell-\frac{1}{2}} + \sin(\alpha)\hat\tau^z_{\ell-\frac{1}{2}}$,  so $\hat T_{\ell+\frac{1}{2}}(\beta) = \tilde\tau^x_{\ell-\frac{1}{2}}$ and $\hat{G}_\ell (\beta)=\hat{G}_\ell$ is the emergent $\mathbb{Z}_2$ gauge generator defined in Eq.~(\ref{Gl}).

	If the specific eigenstate has emergent $\mathbb{Z}_2$ gauge invariance and the initial state has a large overlap with this state,
	then the expectation value of $\hat{G}_\ell$ during the quench dynamics will be nearly time-independent.
	Furthermore, the steady expectation value of $\hat{G}_\ell (\alpha)$ during the dynamics approaches the minimum/maximum when $\alpha=\beta$~\cite{SM}.
	Therefore, based on these considerations, we can determine $\beta$ and thus fix $\hat{G}_\ell$ in the experiment.
	To measure the expectation value of $\hat G_\ell(\alpha)$,
	we need the joint readout of adjacent three qubits under the specific basis, which is accessible in superconducting circuits.
	We should measure the expectation values of $\hat\tau^z_{\ell-\frac{1}{2}} \hat{s}_\ell^z\hat\tau^z_{\ell+\frac{1}{2}} $, $\hat\tau^z_{j-\frac{1}{2}} \hat{s}_\ell^z\hat\tau^x_{\ell+\frac{1}{2}} $, $\hat\tau^x_{\ell-\frac{1}{2}} \hat{s}_\ell^z\hat\tau^z_{\ell+\frac{1}{2}} $,
	and $\hat\tau^x_{\ell-\frac{1}{2}} \hat{s}_\ell^z\hat\tau^x_{\ell+\frac{1}{2}} $, respectively.
	Then, we  combine these four values  linearly according  to Eq.~(\ref{Qj}).

	In Fig.~\ref{fig_4}(a), we show the relation between $\braket{\hat G_3}_\infty$ and $\alpha$ when  $h_x/2\pi=6$~MHz, $V_{2\ell}/2\pi=15$~MHz, and  $\theta = -\pi/3$.
	Here, $\braket{\hat G_3}_\infty$ is the average of $\braket{\hat G_3}$ during the last 0.8~$\mu$s representing the steady value of $\braket{\hat G_3}$.
	We find that $\braket{\hat G_3}_\infty$ is $\alpha$-dependent,
	and it can approach the minimum when $\alpha=\beta\approx-1.19$.
	According  to Fig.~\ref{fig_4}(b), we can find that $\braket{\hat G_3(-1.19)}$ is stabilized at a nonzero value (The oscillation is from the high-order term and absent for effective Hamiltonian $\hat H_{\text{eff}}$~\cite{SM}.). 
	This result provides  evidence that some specific eigenstates of $\hat H_{\text{eff}} $ are almost the eigenstates of $\hat G_\ell$.
	That is, these eigenstates of the effective Hamiltonian can emerge a $\mathbb{Z}_2$ gauge invariance.

	According to $\hat H_{\text{eff}}$, when the external field $\sqrt{h_x^2+h_z^2}$ is much larger than the coupling strength,
	the $\tau$ sector will become the leading contribution of the energy, and the initial state is much close to the ground state when $\theta = -\pi/2-\beta$.
	From Figs.~\ref{fig_3}(b) and ~\ref{fig_4}(a), we can find that $\theta_m$ almost equals $-\pi/2-\beta$,
	and the localization of the $s$-sector mainly exists at the vicinity of $\theta_m$, i.e., in low-energy regimes.
	Therefore, we conjecture that the $\mathbb{Z}_2$ gauge invariance  mainly emerges in the low-energy states of $\hat H_{\text{eff}}$.
	In the SM~\cite{SM}, by obtaining wave functions of all eigenstates of $\hat H_{\text{eff}}$ by exact diagonalization method,
	we can find that the gauge invariance indeed mainly emerges in  low-energy regimes.

	\textit{Summary.}---%
	In conclusion, we have experimentally investigated the emergent $\mathbb{Z}_2$ gauge invariance in a 10-qubit superconducting processor.
	Our experimental results demonstrate that $\mathbb{Z}_2$ gauge invariance can indeed emerge in a low-energy regime,
	even though the $\mathbb{Z}_2$ gauge structure is absent in the effective Hamiltonian.
	Moreover, this emergent gauge invariance can lead to exotic dynamical behaviors, for instance, confinement-induced localization,
	which has been observed in this experiment.
	Our results can scale up to larger quantum systems and enable the further study of emergent  LGTs in superconducting circuits.
	For instance, the dynamics of string breaking~\cite{Hebenstreit2013}, the thermalization of the effective Hamiltonian $\hat H_{\text{eff}}$ and whether existing disorder-free many-body localization~\cite{Smith2017,Brenes2018} are interesting issues.
	In addition, how to realize a truly gauge-invariant Hamiltonian on superconducting circuits is an another relevant question.

	\begin{acknowledgements}
		This work was  supported by 
		the State Key Development Program for Basic Research of China (Grant No. 2017YFA0304300),
		the Key-Area Research and Development Program of Guangdong Province, China (Grant No. 2020B0303030001)
		the National Natural Science Foundation of China (Grant Nos. T2121001,  and 11934018), 		
		the Strategic Priority Research Program of Chinese Academy of Sciences (Grant No. XDB28000000),
		Scientific Instrument Developing Project of Chinese Academy of Sciences (Grant No.~YJKYYQ20200041),
		Beijing Natural Science  Foundation (Grant No.~Z200009),.
	\end{acknowledgements}


	%

\clearpage \widetext

\begin{center}
	\large{\textbf{Supplemental Material}}
	\\ 
	\large{\textbf{\textit{Observation of Emergent  $\mathbb{Z}_2$  Gauge Invariance in a Superconducting Circuit}}}
\end{center}
\setcounter{equation}{0} \setcounter{figure}{0}
\setcounter{table}{0} \setcounter{page}{1} \setcounter{secnumdepth}{3} \makeatletter
\renewcommand{\theequation}{S\arabic{equation}}
\renewcommand{\thefigure}{S\arabic{figure}}
\renewcommand{\thetable}{S\arabic{table}}
\renewcommand{\bibnumfmt}[1]{[S#1]}
\renewcommand{\citenumfont}[1]{S#1}

\makeatletter
\def\@hangfrom@section#1#2#3{\@hangfrom{#1#2#3}}
\makeatother

\maketitle
In this Supplemental Material, we mainly present more details about experiments, 
including information of the device, corrections  of pulses, and other extended experimental and numerical data.

\section{ Experimental Device}

\subsection{Parameters of the chip}
The chip of this experiment is a chain-like superconducting circuit consisting of 10 Xmon qubits (from left to right of the chain is  $Q_1$ to $Q_{10}$, see Fig. 1(a) in the main text),
which are fabricated on a 10~mm$\times$10~mm$\times$0.43~mm sapphire substrate with two steps of aluminum deposition~\cite{Guoxy2021}. 

The detailed parameters of the device are listed in Tab.~\ref{tab1}.
The frequency of readout resonator $f_{r}$ increases by about 20~MHz from $Q_1$ to $Q_{10}$
distributed from 6.51 GHz to  6.69 GHz.
Each Xmon qubit can reach its maximum frequency  $f_{m}$ at the sweetpoint, at which the qubit is
insensitive to flux noise and exhibits long dephasing time.
In this experiment, all qubits are initialized to the ground
state ($\ket{0}$) at their idle frequencies $f_{i}$, which are in the range from 4.934~GHz to 5.692~GHz. 
The single-qubit gate used in the experiment is also performed at their idle frequencies.
Additionally, in this experiment, the working points of odd and even qubits are about 5.14 GHz and 5.22 GH, respectively.
Due to the different capacitors between odd and even qubits, anharmonicity
$\eta$ of the qubits is staggered.
Here, the energy relaxation time (decoherence time) $T_{1}$ and the dephasing time $T_{2}^{*}$ are both measured at the idle point.
We optimize the quadrature correction term with DRAG coefficient $\alpha$  to minimize leakage to higher levels~\cite{2009Simple}. 
Finally, we use the randomized benchmark (RB) method to characterize the  error  of $X/2$ and $Y/2$ gate.
The nearest neighbor (NN) coupling strength is about 12~MHz, and the next nearest neighbor (NNN) coupling
strength  between the odd qubits and between the even qubits are different
due to the different effective capacitance of even and odd qubits.

\begin{table*}[tbh]
	\vspace{15pt}
	\setlength{\tabcolsep}{7pt}
	\centering
	\resizebox{\textwidth}{!}{
		\begin{tabular}{c |c c c c c c c c c c  }
			\hline
			\hline
			& $Q_1$ & $Q_2$ & $Q_3$ & $Q_4$ & $Q_5$ & $Q_6$ & $Q_7$ & $Q_8$ & $Q_9$ & $Q_{10}$ \\
			\hline
			$f_{r}$(GHz)  & 6.514 & 6.536 & 6.554 & 6.575 & 6.596 & 6.614 & 6.634 & 6.651 & 6.671 & 6.689 \\
			$f_{m}$(GHz)  & 5.223 & 5.553 & 5.207 & 5.593 & 5.156 & 5.695 & 5.181 & 5.660 & 5.142 & 5.656 \\
			$f_{i}$(GHz) & 5.037 & 5.534 & 4.934 & 5.543 & 5.064 & 5.692 & 5.00 & 5.468 & 5.113 & 5.570 \\
			$\eta$(MHz)  & -204 & -253 & -206 & -253 & -204 & -249 & -207 & -251 & -205 & -247 \\
			$T_{1}(us)$  & 31.8 & 32.8 & 34.8 & 32.2 & 38.9 & 25.9 & 37.0 & 18.8 & 25.5 & 35.5 \\
			$T_{2}^{*}(us)$  & 1.688 & 3.326 & 0.685 & 1.83 & 4.539 & 2.034 & 1.242 & 2.197 & 5.657 & 3.809 \\
			$F_{ej}$  & 0.897 & 0.874 & 0.865  & 0.916 & 0.880 & 0.932 & 0.908 & 0.845& 0.888 & 0.893 \\
			$F_{gj}$  & 0.970 & 0.958 & 0.966  & 0.989 & 0.959 & 0.984 & 0.976 & 0.968& 0.956 & 0.979 \\
			X/2 Error(\%)  & 0.15 & 0.71 & 0.25  & 0.62 & 0.15 & 0.30 & 0.20 & 0.20& 0.05 & 0.30 \\
			Y/2 Error(\%)   & 0.20 & 0.51 & 0.25  & 0.56 & 0.40 & 0.30 & 0.35 & 0.35& 0.10 & 0.30 \\
			\hline
			\hline
			&$Q_{1-2}$ & $Q_{2-3}$ & $Q_{3-4}$ & $Q_{4-5}$ & $Q_{5-6}$ & $Q_{6-7}$ & $Q_{7-8}$ & $Q_{8-9}$ & $Q_{9-10}$& - \\
			\hline
			$g_{j,j+1}$(MHz)  & 12.05 & 12.2 & 11.90 & 11.90 & 11.90 & 11.76 & 11.90 & 12.05 & 12.35 & -\\      
			\hline
			&$Q_{1-3}$ & $Q_{2-4}$& $Q_{3-5}$ & $Q_{4-6}$ & $Q_{5-7}$ & $Q_{6-8}$ & $Q_{7-9}$ & $Q_{8-10}$ & -& - \\
			\hline
			$g_{j,j+2}$(MHz)  & 1.10 & 0.69 & 1.10 & 0.69 & 1.10 & 0.61 & 1.10 & 0.71 & -& -\\
			\hline
	\end{tabular}}
	\caption{Basic device parameters. $f_{r}$ is the readout resonator
		frequency, $f_{m}$ is the qubit maximum frequency, and $f_{i}$ is
		the qubit idle frequency. $\eta$ is the qubit anharmonicity. $T_{1}$ and $T_2^*$
		are the energy relaxation time and dephasing time of the qubit at idle point.
		$F_{gj}$ and $F_{ej}$  are the readout fidelities for the ground
		and first-excited states, respectively. The  errors of $X/2$ and $Y/2$ gates are also presented.
		In addition, $g_{j,j+1}$ and $g_{j,j+2}$ are the
		coupling strengths of nearest-neighbor (NN) and next-nearest-neighbor (NNN)
		qubits, respectively.}
	\label{tab1}
\end{table*}

\subsection{ Experimental  setup}
\begin{figure*}[htb]
	\centering
	\includegraphics[width=0.9\textwidth]{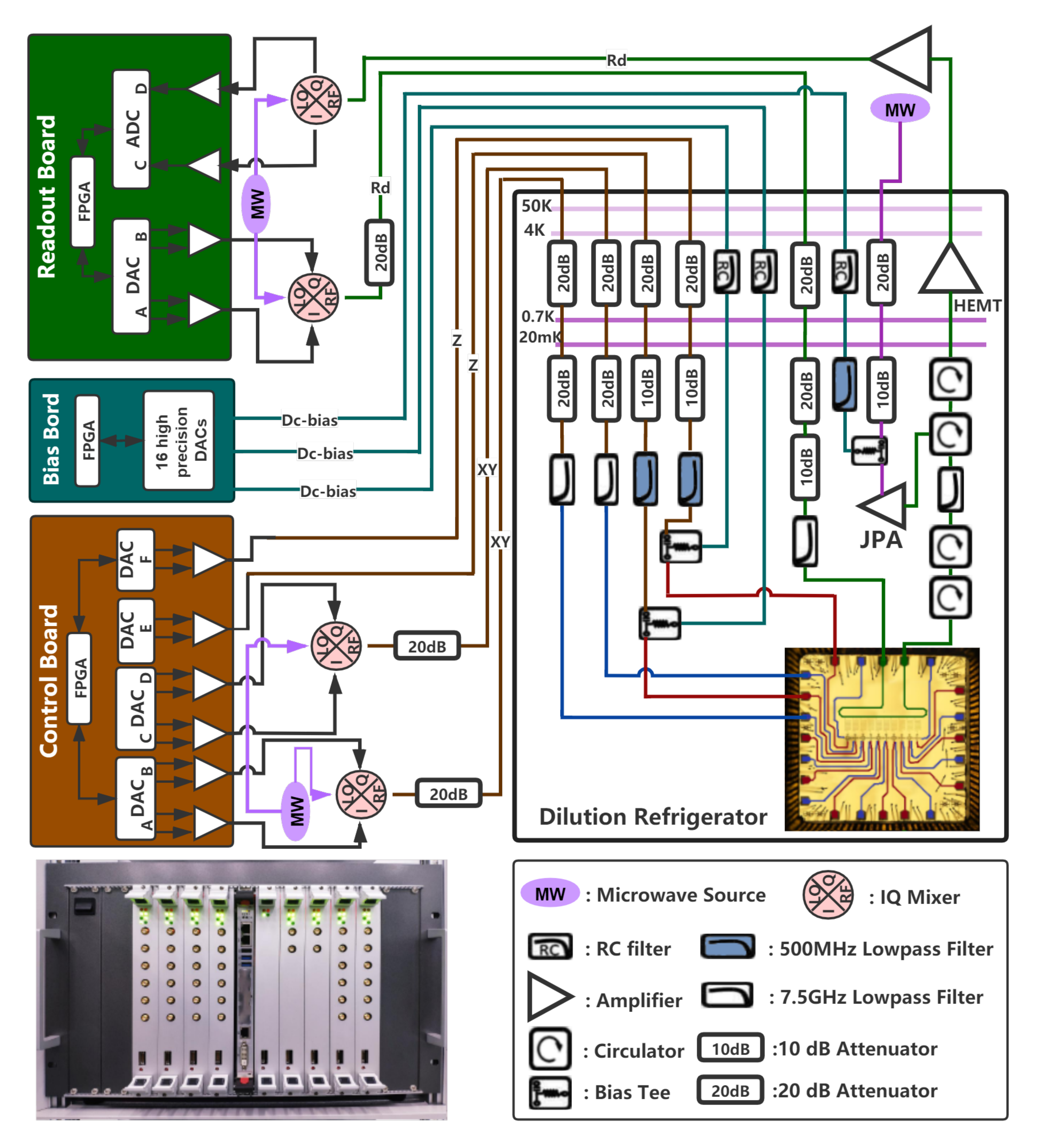}  	
	\caption{Diagram of the experimental setup. 
		The left bottom  is the electronic control instrument system. }	
	\label{fig_s1}
\end{figure*}

In Fig.~\ref{fig_s1}, we present the diagram of the experimental setup. 
This system is consisted of some main function boards and some auxiliary boards. 
The main function boards, displayed at the left top, includes control board, bias board, and readout board. 
The control board is composed of 6 DACs controlled by FPGA, and is used to realize the full XY control  and partial Z control. 
DACs together with the nearby microwave source output microwaves for the XY control of each qubits.
The dc-bias wires compose the bias board, which is used to complete Z control together with the Z wire in control board and bias tee.
The readout board provides the measurement function of qubits.  
The DAC in the readout board and the nearby microwave source output a ten-tone microwave pulse targeting all  readout resonators.
The readout signal is amplified sequentially by the Josephson parametric amplifier (JPA) (JPA is unused in this experiment), high electron mobility transistor (HEMT),
and room temperature amplifiers before demodulated by the ADC.  All control lines go through various stages of attenuation and filter to
prevent unwanted noises from disturbing the operation of the device.

\section{ CALIBRATION}
\subsection{Readout calibration}
\begin{figure*}[ht]
	\centering
	\includegraphics[width=0.9\textwidth]{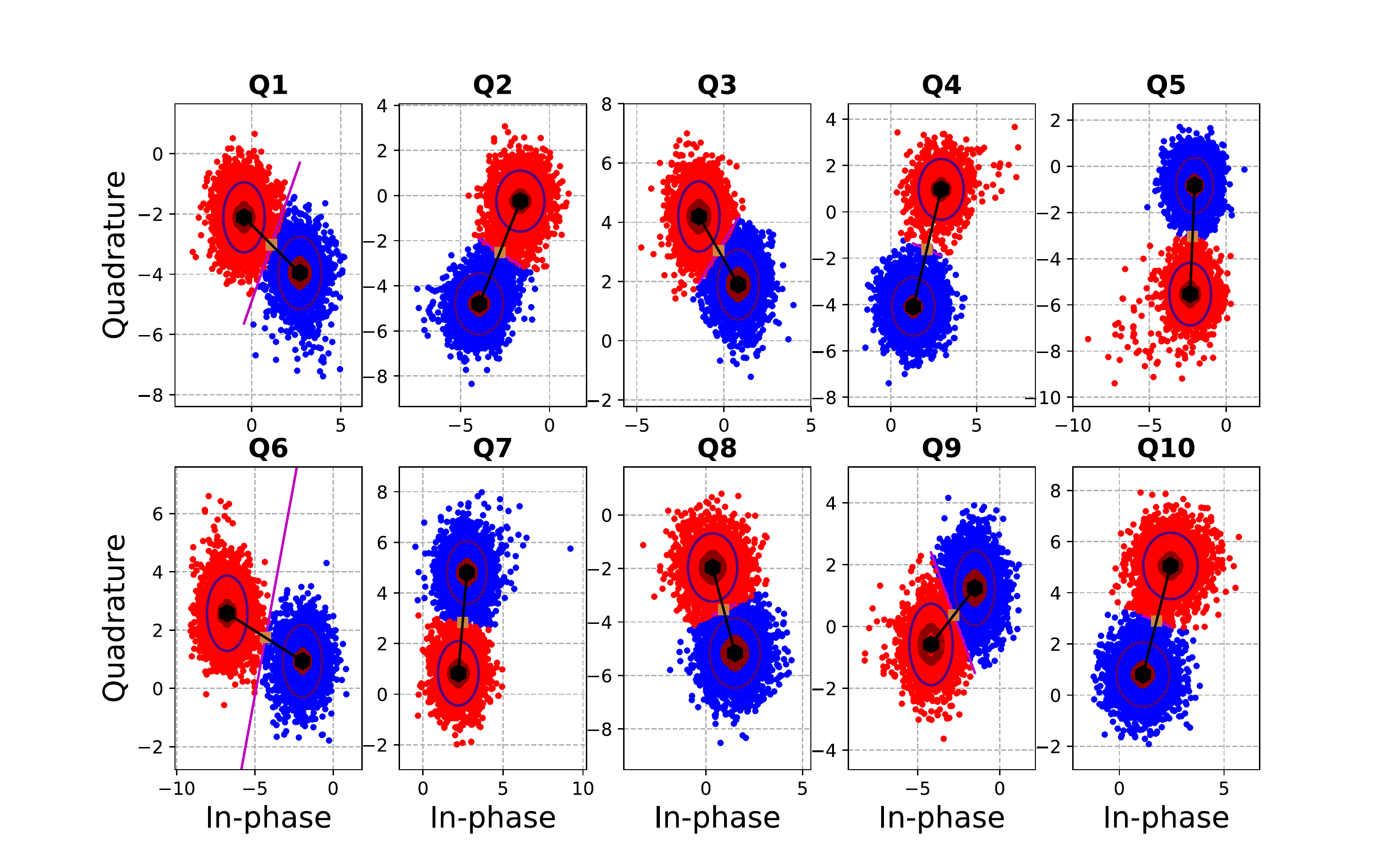}  
	\caption{10 qubit Readout. The red dots means the measurement state is in $\vert 1 \rangle$ and the bule dots means the measurement state is in $\vert 0\rangle$}
	\label{fig_s2}
\end{figure*}

The qubit readout pulse consists of a 1.6~$\mu$s microwave pulse,
which contains information of all readout resonance.
After demodulation by FPGA, we can obtain IQ data, see Fig.~\ref{fig_s2}. 
Due to the unwanted noise, readout has errors. 
Here, we can use the calibration matrix  to calibrate this error, which reads
\begin{equation}
	\left(
	\begin{array}{cc}
		P_g^{\text{c}}        \\
		P_e^{\text{c}}   \\
	\end{array}
	\right)=
	\left(
	\begin{array}{cc}
		F_{gj} & 1-F_{ej}\\
		1-F_{gj} & F_{ej}\\          
	\end{array}
	\right)
	\left(
	\begin{array}{cc}
		P_g       \\
		P_e   \\
	\end{array}
	\right),
\end{equation}
where $F_{gj}$ and $F_{ej}$ are readout fidelities of state $\ket{0}$ and $\ket{1}$ of $Q_j$, respectively, see Tab.~\ref{tab1}, 
and $P_g^{\text{c}}$ ($P_e^{\text{c}}$) and $P_g$ ($P_e$) are calibrated and original probabilities 
(calculated directly from IQ data) of  $\ket{0}$ ($\ket{1}$), respectively.

\subsection{Crosstalk correction of  Z pulse }
The  crosstalk of Z lines will decrease the accuracy of the experiment, and thus should be corrected.
Firstly, we need determine the Z crosstalk matrix $M_z$, which can be calculated by measuring the offset and 
the compensation offset of the qubit through different Z lines. The measured Z crosstalk matrix  in this system is shown in Tab.~\ref{tab2}.
Here, if we applied the Z pulse to each qubit with strength $Z_{\text{app}}$, then the qubits can actually  feel  the strength  $Z_{\text{act}} = M_{z} \cdot Z_{\text{app}}$.
Thus, we can get the compensation value through the above formula to calibrate the error caused from Z crosstalk.

\begin{table*}[tbh]
	\vspace{15pt}
	\setlength{\tabcolsep}{7pt}
	\centering
	\resizebox{\textwidth}{!}{
		\begin{tabular}{c |c c c c c c c c c c  }
			\hline
			& $Q_1$ & $Q_2$ & $Q_3$ & $Q_4$ & $Q_5$ & $Q_6$ & $Q_7$ & $Q_8$ & $Q_9$ & $Q_{10}$ \\
			\hline
			$Q_1$  & 1.0000 & -0.0032 & 0.0112 &  0.0031 & -0.0012 & 0.0064 & 0.0033 & 0.0037 & 0.0027 &0.0034\\
			$Q_2$  & -0.0023&  1.0000&  -0.0041& -0.0033& 0.0000&-0.0036&  -0.0020&  -0.0026& -0.0021 & -0.0024 \\
			$Q_3$  &-0.0058&  -0.0377&  1.0000&  -0.0077& 0.0005& -0.0075& -0.0049&  -0.0065& -0.0052& -0.0057 \\
			$Q_4$  & -0.0028&  -0.0068& -0.0085&  1.0000& -0.0029& -0.0026&  0.0031& -0.0027& -0.0028&  -0.0028 \\
			$Q_5$  & 0.0051&  0.0102&  0.0081&  0.0297&  1.0000&0.0039&    0.0050&   0.0069&  0.0055& 0.0049 \\
			$Q_6$  & 0.0033&  0.0061&  0.0040& 0.0090& -0.0102& 1.0000&  0.0082&  0.0064&  0.0045& 0.0037\\
			$Q_7$  & -0.0049& -0.0085& -0.0050& -0.0098&  0.0060& 0.0188&  1.0000&  -0.0220& -0.0103&  -0.0061\\
			$Q_8$  & -0.0030& -0.0049&  -0.0028& -0.0053& 0.0028&0.0052& 0.0010&   1.0000& -0.0113& -0.0042 \\
			$Q_{10}$ &0.0031&  0.0045&  0.0024&  0.0048&  -0.0026&-0.0053& -0.0025&  -0.0075& -0.0072&  1.0000 \\
			\hline
	\end{tabular}}
	\caption{Z crosstalk matrix $M_z$.}
	\label{tab2}
\end{table*}

\subsection{Distortion calibration of Z pulse }
\begin{figure*}[ht]
	\centering
	\includegraphics[width=0.5\textwidth]{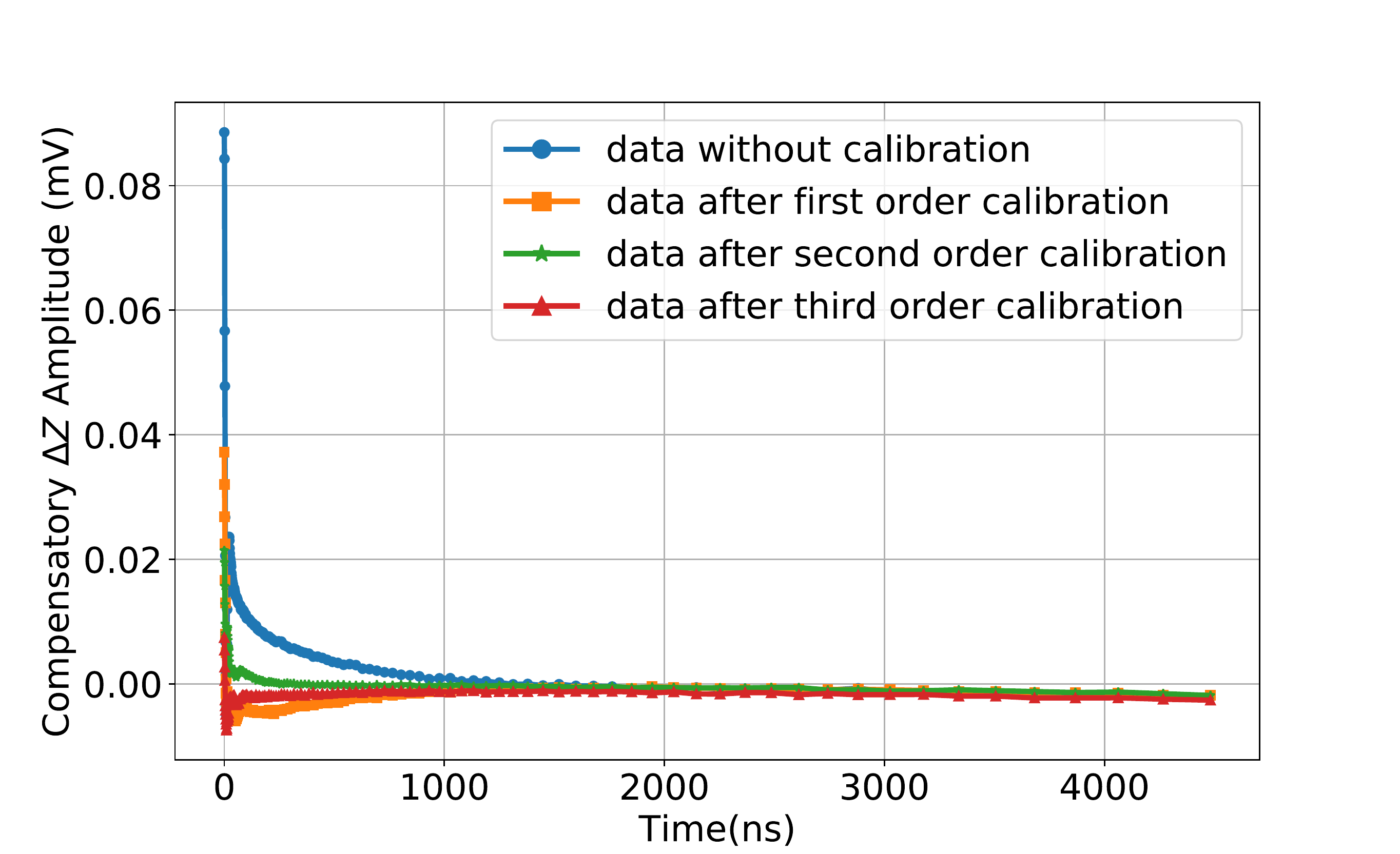}
	\caption{Z pulse calibration.}
	\label{fig_s3}
\end{figure*}

In the experiment, we need adjust the qubit level  by applying a square wave pulse to the Z line. 
However, due to the presence of parasitic inductance and capacitance, an ideal square pulse is usually distorted when reaching to the chip showing overshoot or undershoot near the rising edge and tailed falling edge.
In Fig.~\ref{fig_s3}, we show the results for distortion calibration of Z pulse.
We can find that, after third order calibration, Z pulse can nearly become a perfect square wave.

\section{Low-energy physics of the effective Hamiltonian}
In this section, we use matrix product state (MPS) based methods
to study the ground state and quench dynamics properties of the effective Hamiltonian, i.e., Eq.~(2) in the main text.
Without loss generality, we consider a homogeneous system, namely, we fix $ \tilde{g}_{s/\tau,\ell}= g=1.8$, 
$\lambda_{2\ell-1}=\lambda_s=1.1$, and $\lambda_{2\ell}=\lambda_\tau=0.7$.
Thus, the Hamiltonian  reads
\begin{align} \label{Heff}
	& \hat H_{\text{eff}}=\hat H_1 + \hat H_2+\hat H_3, \\ \nonumber
	&\hat  H_1 =-\sum_{\ell=1}g(\hat{s}^+_{\ell}\hat{\tau}^z_{\ell+\frac{1}{2}}\hat{s}^-_{\ell+1} +
	\text{H.c.} ) +\sum_{\ell=1}h_x\hat{\tau}_{\ell+\frac{1}{2}}^x  ,\\ \nonumber
	& \hat H_2 = \sum_{\ell=1}g(\hat{\tau}^+_{\ell-\frac{1}{2}} \hat{s}^z_{\ell}\hat{\tau}^-_{\ell+\frac{1}{2}}+ \text{H.c.}),\\ \nonumber
	& \hat H_3 \!= \!\sum_{\ell=1}(\lambda_{s}\hat{s}^+_{\ell} \hat{s}^-_{\ell+1} \!+\!\lambda_{\tau}\hat{\tau}^+_{\ell+\frac{1}{2}} \hat{\tau}^-_{\ell+\frac{3}{2} }+
	\text{H.c.} ) \!+\!\sum_{\ell=1}h_z\hat{\tau}_{\ell+\frac{1}{2}}^z.
\end{align}
Thus, there are only two driving parameters, i.e., $h_x$ and $h_z$.
Here, we note that $h_z$ has two parts: one is original longitudinal fields of $\tau$-spins (even qubits), i.e., $V_{2\ell}$ in Eq.(1) of main text,
and the other is the high-order correction from Schrieffer-Wolf transformation, i.e., Lamb shift. 
In Ref.~\cite{Ge2020}, $h_z$ is fixed to $0$ by choosing a proper $V_{2\ell}$, but we let $h_z$ as a driving parameter in this experiment.
The Lamb shift can hardly be confirmed accurately, since we cannot calculate all orders of Schrieffer-Wolf transformation.
However, $V_{2\ell}$ is adjustable in the experiment, which is the detuning between the corresponding qubit frequencies and XY driving frequencies.

Here, we can find that $\hat H_1$ is a typical $\mathbb{Z}_2$ LGT coupled with a matter field~\cite{Schweizer2019,Borla2020}, where $\tau$ and $s$  are gauge and matter fields, respectively, and the transverse field $h_x$ is the corresponding $\mathbb{Z}_2$ electric field.
The  $\mathbb{Z}_2$ gauge transformation can be defined as $\hat{G}^0_\ell = \hat{\tau}_{j-\frac{1}{2}}^x\hat{s}_j^z\hat{\tau}_{j+\frac{1}{2}}^x$ satisfying $[\hat{G}^0_\ell, \hat H_1]=0$.
However, $\hat H_2$ and  $\hat H_3$ both lack this $\mathbb{Z}_2$  invariance, i.e.,
$[\hat{G}^0_\ell, \hat H_2]\neq0$ and $[\hat{G}^0_\ell, \hat H_3]\neq0$
Therefore, the whole effective Hamiltonian $\hat H_{\text{eff}}$ is not $\mathbb{Z}_2$ gauge invariant.

To analyze the emergent gauge invariance of $\hat H_{\text{eff}}$, we map the $\tau$-spin (gauge field) to another frame, i.e., performing the following replacement
\begin{align} \label{tau}
	& \tilde\tau^x_{\ell+\frac{1}{2}}=\cos \beta \hat\tau^x_{\ell+\frac{1}{2}} + \sin\beta\hat\tau^z_{\ell+\frac{1}{2}},\\ \nonumber
	& \tilde\tau^y_{\ell+\frac{1}{2}}=\hat\tau^y_{\ell+\frac{1}{2}},\\ \nonumber
	&\tilde\tau^z_{\ell+\frac{1}{2}}=\cos \beta \hat\tau^z_{\ell+\frac{1}{2}} - \sin\beta\hat\tau^x_{\ell+\frac{1}{2}},
\end{align}
where $\beta= \arctan (h_x/h_z)$.
Thus, the effective Hamiltonian $\hat H_{\text{eff}}$ can be transformed as
\begin{align} \label{Heff2}
	& \hat H_{\text{eff}}=\tilde H_1 + \tilde H_2 + \tilde H_3, \\ \nonumber
	&\tilde  H_1 =-\sum_{\ell=1}g\cos \beta (\hat{s}^+_{\ell}\tilde{\tau}^z_{\ell+\frac{1}{2}}\hat{s}^-_{\ell+1} +
	\text{H.c.} ) +\sum_{\ell=1}\tilde h\tilde{\tau}_{\ell+\frac{1}{2}}^x  ,\\ \nonumber
	& \tilde H_2 =\sum_{\ell=1}\hat{s}^+_{\ell}( \lambda_{s}-g\sin \beta \tilde\tau^x_{\ell+\frac{1}{2}} )\hat{s}^-_{\ell+1}+
	\text{H.c.} ,\\ \nonumber
	& \tilde H_3 =\frac{g}{2}\sum_{\ell=1}[\tilde{\tau}^y_{\ell-\frac{1}{2}}(\lambda_{\tau}- \hat{s}^z_{\ell})\tilde{\tau}^y_{\ell+\frac{1}{2}}
	\!+\!\cos^2\!\beta \tilde{\tau}^x_{\ell-\frac{1}{2}}(\lambda_{\tau}- \hat{s}^z_{\ell})\tilde{\tau}^x_{\ell+\frac{1}{2}}\\ \nonumber
	& \!+\!\sin^2\!\beta \tilde{\tau}^z_{\ell-\frac{1}{2}}(\lambda_{\tau}- \hat{s}^z_{\ell})\tilde{\tau}^z_{\ell+\frac{1}{2}}
	\!+\!\sin\!\beta \cos\!\beta\tilde{\tau}^x_{\ell-\frac{1}{2}}(\lambda_{\tau}- \hat{s}^z_{\ell})\tilde{\tau}^z_{\ell+\frac{1}{2}}\\ \nonumber
	&	+\sin\beta \cos\beta\tilde{\tau}^z_{\ell-\frac{1}{2}}(\lambda_{\tau}- \hat{s}^z_{\ell})\tilde{\tau}^x_{\ell+\frac{1}{2}}],
\end{align}
where $\tilde h=\sqrt{h_x^2+h_z^2}$.
Similarly, we can find that $\tilde H_1$ is also a  $\mathbb{Z}_2$ LGT with gauge generator 
\begin{align} \label{Gl}
	\hat{G}_\ell= {\tilde\tau}_{\ell-\frac{1}{2}}^x\hat{s}_\ell^z{\tilde\tau}_{\ell+\frac{1}{2}}^x.
\end{align}
However, $\tilde H_2$ and  $\tilde H_3$ violate such $\mathbb{Z}_2$ gauge invariance.

\begin{figure*}[t]     		
	\includegraphics[width=0.6\textwidth]{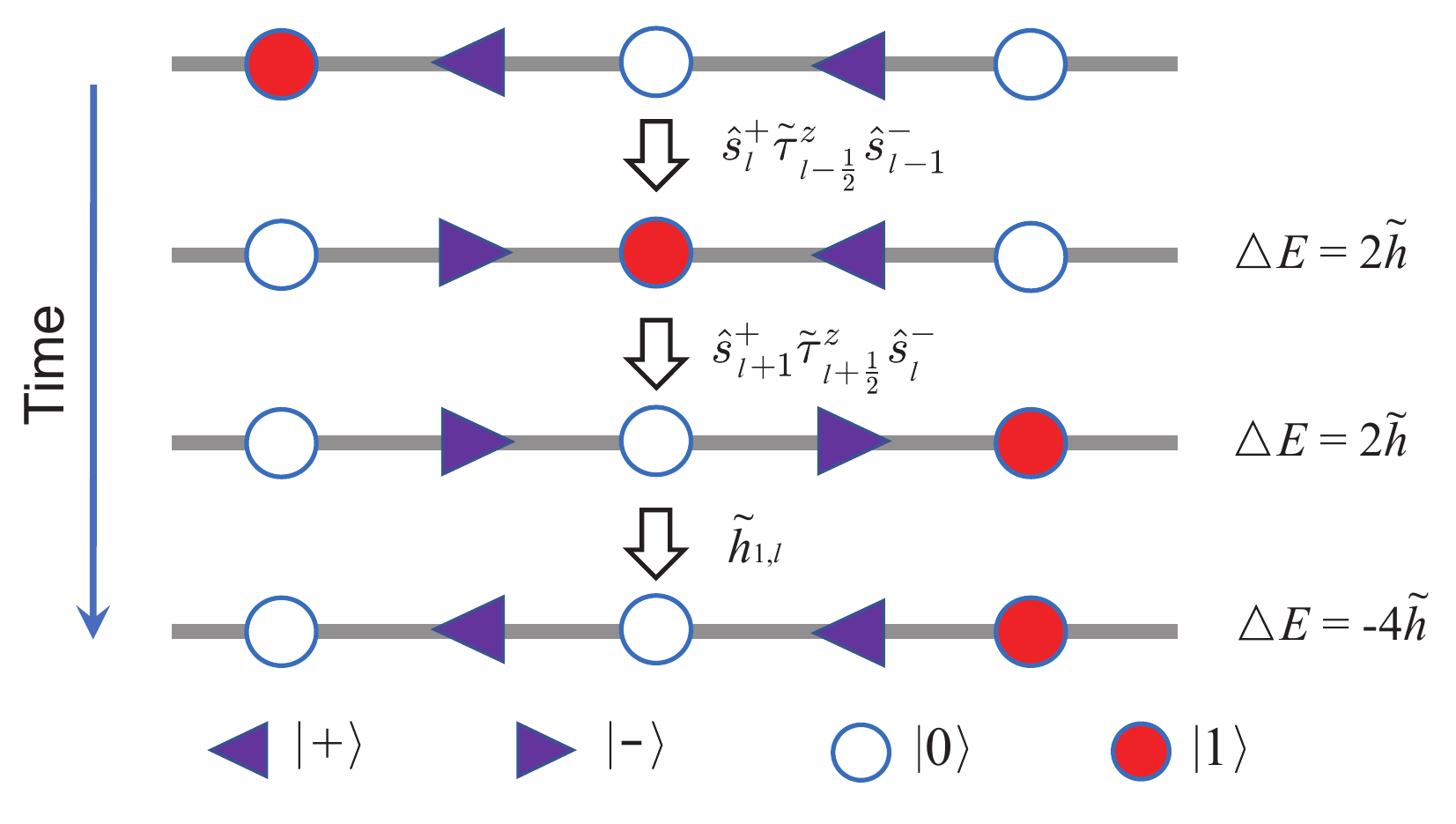}
	\caption{Diagram of the action of $\tilde  H_1$ and $\tilde h_{1,\ell}$.} 
	\label{fr1}
\end{figure*}

Now we discuss the ground state of $ \hat H_{\text{eff}}$.
When $\tilde h\gg g$, $\tilde \tau$ sector is almost polarized in the $\tilde \tau^x$ channel under the ground state.
Thus, $\braket{\tilde{\tau}_{\ell+\frac{1}{2}}^x}\neq0$, where $\langle \cdot\rangle$ represents taking expectation value towards the ground state.
Then, using mean-field approximation, 
\begin{align} \label{H2}
	\tilde H_2\approx\sum_{\ell=1}\hat{s}^+_{\ell}( \lambda_{s}-g\sin \beta \braket{\tilde{\tau}_{\ell+\frac{1}{2}}^x})\hat{s}^-_{\ell+1}+
	\text{H.c.}.
\end{align}
Under the specific fields $h_x$ and $h_z$, we can have $\lambda_{s}-g\sin \beta \braket{\tilde{\tau}_{\ell+\frac{1}{2}}^x}=0$, i.e., $\tilde H_2=0$.
For $\tilde  H_3$, we take the first term, i.e.,  
$\tilde h_{1,\ell} =\tilde{\tau}^y_{\ell-\frac{1}{2}}(\lambda_{\tau}- \hat{s}^z_{\ell})\tilde{\tau}^y_{\ell+\frac{1}{2}}$, 
as an example to show how this term to vanish in terms of low-order perturbations~\cite{Ge2020}.
We can use the following phenomenological description~\cite{Ge2020}: 
As shown in Fig.~\ref{fr1}, under the action of $\tilde  H_1$, the system can firstly have
two imaginary processes to make a $s$-spin hop to the NNN
$s$-site by flipping two $\tau$-spins. Then, these two $\tau$-spins can flip back through the term $\tilde h_{1,\ell}$. 
Here,  the energy after the whole process conserves.
Thus, the term $\tilde h_{1,\ell}$ can contribute at the third order perturbation.
Following this picture, 
we can find that the dominating term of this system is $\tilde H_1$, which is $\mathbb{Z}_2$ gauge invariant.

We also note that, for the small $h_x$ or high initial energy, since $\tau$-spins may not be polarized,
the relation $\lambda_{s}-g\sin \beta \langle {\tilde{\tau}_{\ell+\frac{1}{2}}^x}\rangle=0$ is broken.
Thus, $\tilde  H_2$ cannot be ignored, so the gauge invariance will not exist in these cases.

\begin{figure*}[ht]
	\centering
	\includegraphics[width=0.95\textwidth]{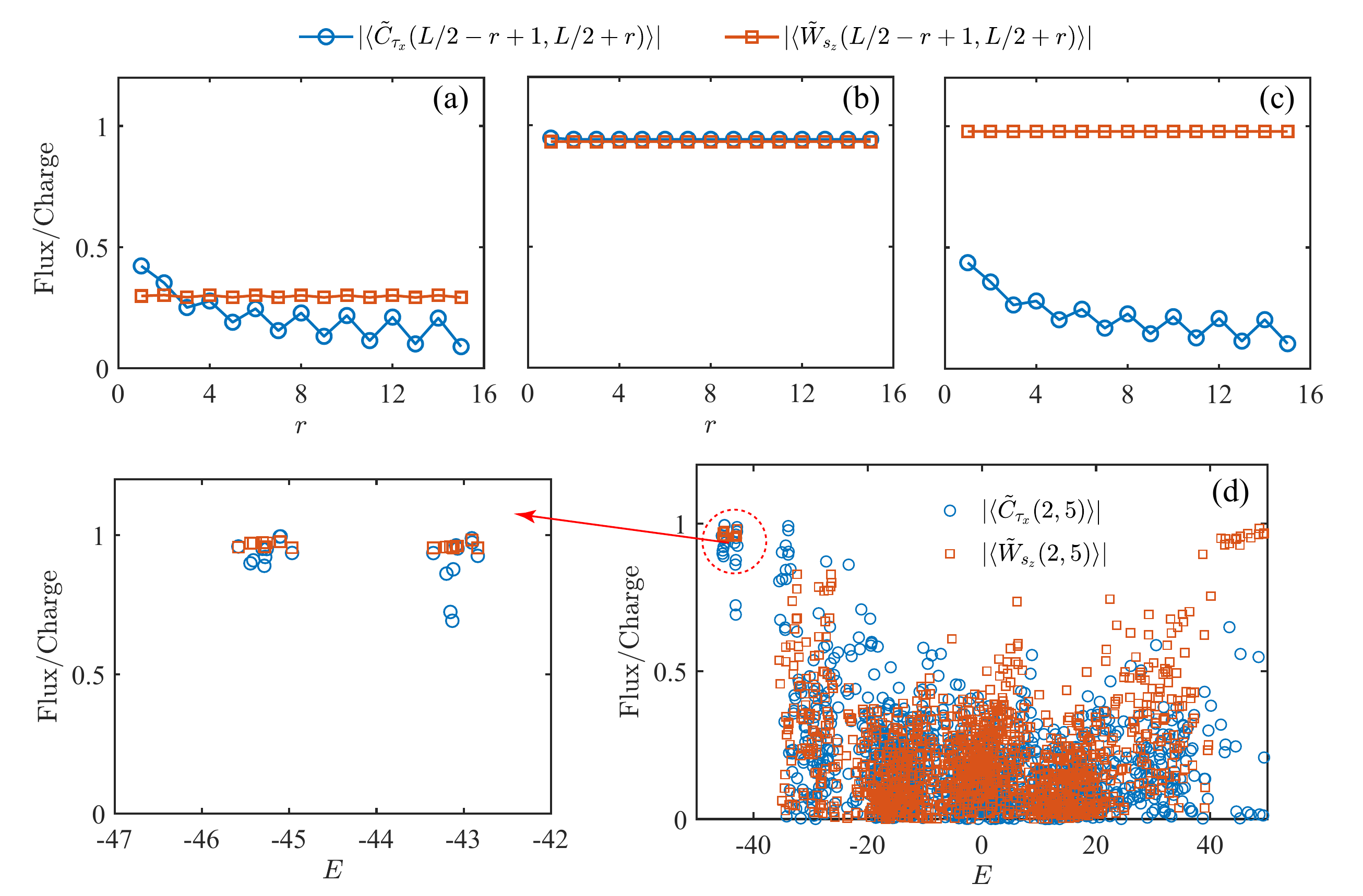}
	\caption{Expectation values of $ \mathbb{Z}_2$ charge and the flux at ground state with $h_x = -6$ and (a) $h_z=0$, (b) $h_z=-4.45$, (c) $h_z=-6$.
		The results are calculated via density matrix renomalized group (DMRG) method~\cite{Schollwock2005,Schollwock2011} with $L=80$. (d) Expectation values of $ \mathbb{Z}_2$ charge and the flux for all eigenstates with  $h_z=-4.45$.
		The result is calculated vis exact diagonalization method with $L=6$. The inset is shows the $ \mathbb{Z}_2$ charge and the flux for low-energy eigenstates. }
	\label{fig_s4}
\end{figure*}

To verify the emergent $\mathbb{Z}_2$ gauge invariance of $ \hat H_{\text{eff}}$ in the ground state, numerically,
we define the $ \mathbb{Z}_2$ charge 
\begin{align} \label{Ws}
	\tilde W_{s_z} (i,j):= \prod_{i\leq k\leq j} \tilde s_k^z,
\end{align}
and the flux
\begin{align} \label{Ct}
	\tilde C_{\tau_x}(i,j) := \tilde \tau_{i-\frac{1}{2}}^x\tilde \tau_{j+\frac{1}{2}}^x.
\end{align}
We know that $\tilde W_{s_z} (i,j)= \pm\tilde C_{\tau_x}(i,j)$ must be  satisfied rigorously for  $\mathbb{Z}_2$ gauge invariant systems (e.g., $\tilde H_1$), which is Gauss law of 1D $\mathbb{Z}_2$ LGT.
Now,  we first use density matrix renomalized group (DMRG) method to calculate $ \mathbb{Z}_2$ charge and flux at the ground sate.
Here, open boundary condition is used for the numerical simulation. 
According to the above picture, we can find that the emergent Gauss's law mainly originates from the polarization of $\tau$-spins.
Thus, the filling of $s$-spins can hardly affect the effective Hamiltonian, 
i.e., the effective Hamiltonian $\tilde{H}_1$ is valid for the arbitrary filling factor. Without loss of generality,
we set the $s$-spins to be half-filling, i.e., $\sum_{j=1}^L \hat{s}_j^z=0$.
From Figs.~\ref{fig_s4}(a-c), one can find, in our system, 
\begin{align} 
	\langle {\tilde C_{\tau_x}(i,j)}\rangle\simeq (-1)^{f(i,j)}\langle {\tilde W_{s_z} (i,j)}\rangle,
\end{align}
when $h_x=6$ and $h_z =- 4.45$, i.e., $\beta\approx-0.638$. 
Here, $f(i,j)=\frac{(j-i+1)(j+i+2)}{2}$ is an integer.
Therefore, the effective Hamiltonian $ \hat H_{\text{eff}}$ can indeed emerge $\mathbb{Z}_2$ gauge invariance at the ground state under proper  transverse and longitude field,
although the $\mathbb{Z}_2$ gauge invariance is absent in $\hat H_{\text{eff}}$.
In Figs.~\ref{fig_s4}(d), we present the expectation values of $\mathbb{Z}_2$ flux and charge for all eigenstates of $ \hat H_{\text{eff}}$ at half-filling regime and periodic boundary condition.
We can find that  $\mathbb{Z}_2$ gauge invariance is almost absent in high excited state. 

Now we use time evolving block decimation  (TEBD) method~\cite{Vidal2004} to study the quench dynamics of $ \hat H_{\text{eff}}$.
Here, we choose second-order Suzuki-Trotter decomposition. 
We also enlarge the maximum bond dimension and decrease time of single step till the final results converge.
We  set the total $s$-charge to $\sum_{j=1}^L \hat{s}_j^+ \hat s_j^- = 1$. The initial state is chosen as $|\psi_0\rangle = |s\rangle\otimes|\tau\rangle$, where  $|s\rangle$ and $|\tau\rangle$ label the states of $s$ and $\tau$ sectors, respectively.
We let $|s\rangle = \ket{...\uparrow\uparrow\downarrow\uparrow\uparrow...}$ and $|\tau\rangle=\ket{...\Phi_\theta\Phi_\theta\Phi_\theta\Phi_\theta\Phi_\theta...}$, where$\ket{\Phi_\theta}=\cos \frac{\theta}{2}\ket{\uparrow}+\sin\frac{\theta}{2}\ket{\downarrow}$.
Thus, this initial state is closely related to the experiment in maintext.
In addition, we know that, when $\theta=-\pi/2-\beta$, the system is much close to the ground state.

\begin{figure*}[ht]
	\centering
	\includegraphics[width=0.95\textwidth]{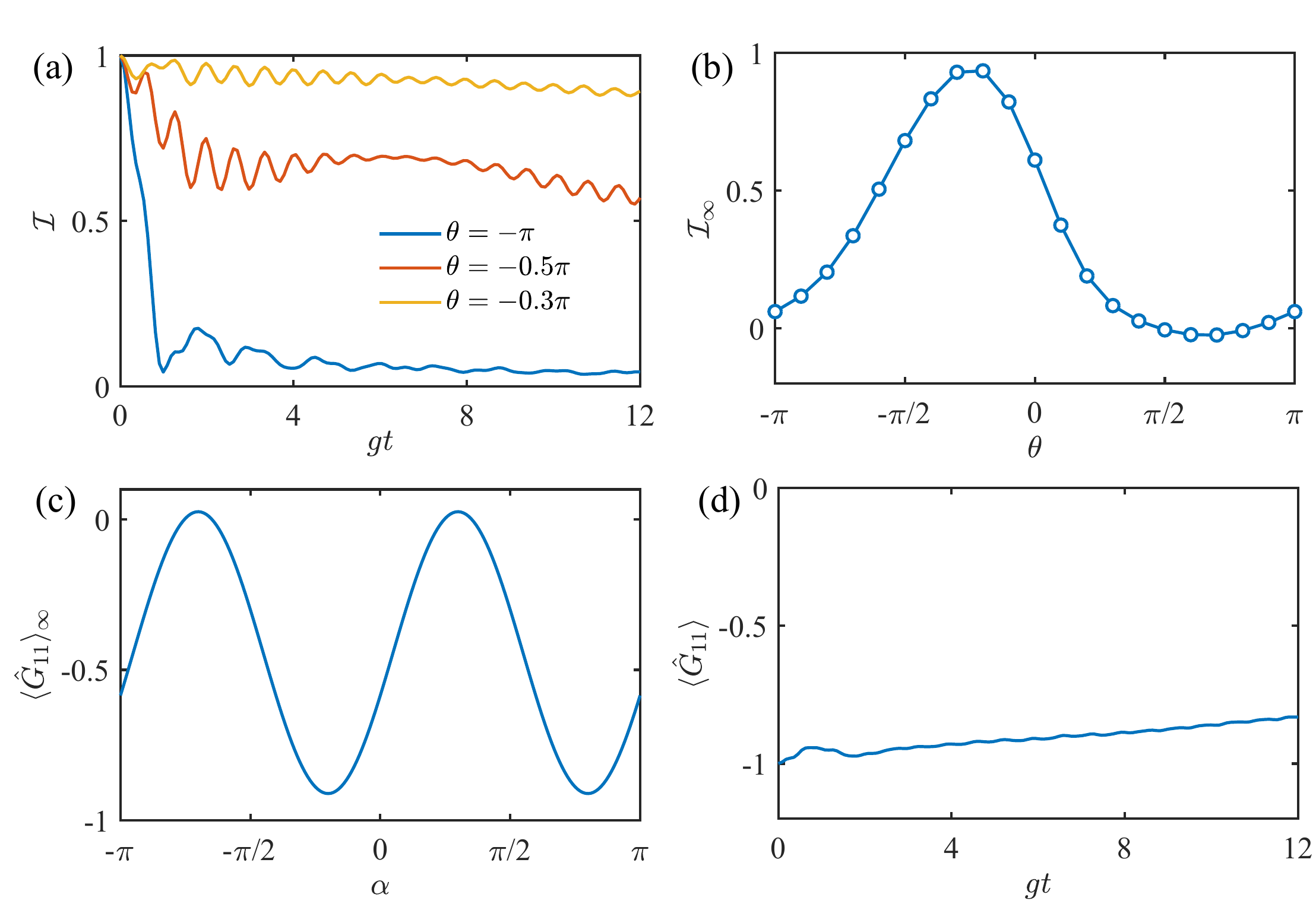}
	\caption{Quench dynamics of $ \hat H_{\text{eff}}$ with $h_x = -6$, $h_z = -4.45$ and $L=21$. 
		(a) Time evolution of  the extended imbalance of $s$-spins for different initial states. 
		(b) The relation between stead values of the extended imbalance $\mathcal{I}_\infty$  and initial states.
		Here, $\mathcal{I}_\infty$ is the average of $\mathcal{I}(t)$ for $t \in [4,8]$.
		(c) The  stead values of $\hat{G}_{11} (\alpha)$, and $\braket{\hat{G}_{11} (\alpha)}_\infty$  is also the average of $\braket{\hat{G}_{11} (\alpha)}(t)$for $t \in [4,8]$.
		(d)  Time evolution of  $\mathbb{Z}_2$ gauge generator  $\hat{G}_\ell(\alpha=\beta=-6.38)$, and there is nearly no oscillation.
		Here, the results are calculated via TEBD method.}
	\label{fig_s5}
\end{figure*}

Firstly, we calculate the time evolution of  the extended imbalance of $s$-spins
defined in Eq.~(5) of main text.
According to Fig.~\ref{fig_s5}(a), we can find $s$-sector can exhibit  a localization in a short time regime when $h_x=6$ and $h_z =- 4.45$,
and localization strength depends on the initial state.
Fig.~\ref{fig_s5}(b) shows that the localization strength of $s$-sector approach the strongest when $\theta\approx-\pi/2-\beta$.
Now we study the  time evolution of $\mathbb{Z}_2$ gauge generator  $\hat{G}_\ell$.
Similar to the main text, we  define an ansatz of $\mathbb{Z}_2$ gauge generator as
\begin{align} \label{Qj}
	\hat{G}_\ell (\alpha) :=\hat T_{\ell-\frac{1}{2}}(\alpha)\hat{s}_\ell^z \hat T_{\ell+\frac{1}{2}}(\alpha),
\end{align}
where $\hat T_{\ell-\frac{1}{2}}(\alpha) = \cos (\alpha)\hat\tau^z_{\ell-\frac{1}{2}} + \sin(\alpha)\hat\tau^x_{\ell-\frac{1}{2}}$,  so $\hat T_{\ell+\frac{1}{2}}(\beta) = \tilde\tau^x_{\ell-\frac{1}{2}}$ and $\hat{G}_\ell (\beta)=\hat{G}_\ell$ is the emergent $\mathbb{Z}_2$ gauge generator.
According to Figs.~\ref{fig_s5}(c--d), we can find that the stead value of $\hat{G}_\ell (\alpha)$ indeed approach the minimum when $\alpha=\beta$.
These quench dynamics is completely consistent with our experimental results.
Here we note that there is nearly no oscillation during the dynamics of $\hat{G}_\ell (\beta)$ for $ \hat H_{\text{eff}}$ [see Fig.~\ref{fig_s5}(d)], which is distinct to the corresponding experimental results [see Fig.~4(b) in main text].
Thus, we conjecture that this oscillation of the dynamics for original Hamiltonian results from the high-order term.



\end{document}